\documentclass[12pt,a4paper]{article}

\usepackage[margin=1in]{geometry}
\usepackage{graphicx}
\usepackage{amsmath, amssymb}
\usepackage{physics}
\usepackage{siunitx}
\usepackage{xcolor}
\usepackage{float}
\usepackage{setspace}
\usepackage{caption}
\usepackage{subcaption}
\usepackage{authblk}
\usepackage[numbers,sort&compress]{natbib}
\usepackage{hyperref}
\usepackage{makecell}

\hypersetup{
    colorlinks = true,
    linkcolor = blue,
    citecolor = blue,
    urlcolor  = blue
}

\title{\textbf{Pristine and Doped MoS\textsubscript{2} Monolayers as Potential HCN Gas Sensors: A DFT Study}}

\author[1,2,*]{Neeraj Thakur}
\author[3]{Anjna Bhardwaj}
\author[4]{Arun Kumar}
\author[1]{Amarjeet Singh}
\affil[1]{Department of Physics, Himachal Pradesh University, Shimla-171005, India}
\affil[2]{School of Mechanical and Materials Engineering, IIT Mandi, Kamand-175075, India}
\affil[3]{Government Degree College Baroh, Kangra-176054, Himachal Pradesh, India}
\affil[4]{Government Degree College Bilaspur, Bilaspur-174001, Himachal Pradesh, India}
\affil[*]{Corresponding author: neerajthak74@gmail.com}

\date{} 

\begin{document}
\maketitle

\begin{abstract}
\noindent
Two-dimensional transition metal dichalcogenides (TMDCs) have been extensively investigated due to their tunable properties. In this work, density functional theory (DFT) was employed to investigate the adsorption behavior and sensing characteristics of HCN on pristine and doped MoS$_2$ monolayers (X–MoS$_2$, where X = P, N, Si, Al, B, Cl). The structural, electronic, and optical characteristics of all systems were examined to study the sensing properties of various doped MoS$_2$ monolayers. In particular, the Al–MoS$_2$ system demonstrated the strongest adsorption characterized by chemisorption, while the remaining systems showed interactions of physisorption type. Recovery time and changes in electronic and optical properties revealed that Si–MoS$_2$ possesses an ultrafast response of the order of microseconds, while Al–MoS$_2$ exhibits a significantly longer recovery time, making it unsuitable for reusable sensors. P–, Si–, and Al–MoS$_2$ monolayers showed pronounced changes in their properties after HCN adsorption. To explore tunability in adsorption strength and recovery behavior, systems with two and three dopant atoms were further studied for P, Si, and Al doping. The results indicate that double doping enhances adsorption strength, whereas triple symmetric doping weakens it. Based on adsorption energy, recovery time, and electronic response, 2P–MoS$_2$ and 3Al–MoS$_2$ are identified as promising candidates for electrochemical and chemiresistive sensing of HCN. Additionally, the observed optical response in the ultraviolet region highlights their potential in UV-range optical sensor design.
\end{abstract}


\clearpage
\section{Introduction}
The sensing and detection of toxic gases is one of the major issues in the industrial process \cite{1,2}. Hydrogen cyanide (HCN) and cyanide ion (CN$^-$) are two highly toxic natural forms of cyanide. Inhalation of these substances can cause serious damage to the human body and may even be fatal. Since these two forms exist in distinct physical states, the detection techniques required for each also differ \cite{2}. With the rapid increase in global demand for development, industrial infrastructure is expanding in almost every sector. This expansion has led to a concerning increase in industrial waste and effluents, particularly the emission of highly toxic gases. To ensure effective environmental monitoring and protect human health, it is essential to detect these pollutants at the molecular level. This necessity has spurred intense research in the area of gas-sensing technologies \cite{1,2,3}. Despite the significant amount of work done on the detection of CN$^-$ ions, targeted investigations concerning HCN detection remain comparatively rare. Moreover, under specific conditions, free cyanide ions are known to convert to HCN in wastewater systems \cite{2}.

Hydrogen cyanide (HCN) is a highly flammable, colorless liquid with a boiling point near ambient temperature (298.75~K) \cite{4}. HCN is recognized for its ability to disrupt cellular respiration by preventing the effective use of oxygen within the body. Notably, HCN has been found in exhaled human breath during pulmonary infections—particularly among cystic fibrosis (CF) patients. The opportunistic bacterium Pseudomonas aeruginosa is known to release HCN, which can be detected in breath samples at concentrations reaching up to several tens of parts per billion (ppb) \cite{2,5}. Additionally, HCN is considered a useful atmospheric tracer for assessing air quality on both local and regional scales \cite{2}. In industrial applications, it functions as a precursor in numerous chemical manufacturing processes \cite{6}.

Consequently, there is a pressing need to develop efficient sensors that exhibit high sensitivity, better recovery times, rapid response, real-time monitoring capabilities, and affordability. Several investigations have examined the interaction of HCN with materials such as graphene, graphyne, carbon nanotubes, boron nitride nanotubes, AlN, and silicene nanoribbons \cite{2,4,6,7,8}. Nevertheless, to the best of our knowledge, the detection of HCN using two-dimensional transition metal dichalcogenides (TMDCs) remains unexplored \cite{2,4,6,7,8}.

Among the TMDC family, the MoS$_2$ \cite{9} monolayer has been extensively studied for sensing various gases, making it a promising and potentially cost-effective candidate for HCN detection. 2D Nano-sheets of MoS$_2$ has already been experimentally synthesized in laboratories for realistic applications \cite{10,11}. However, a single layer exfoliation of MoS$_2$, like graphene, is yet to be achieved. Utilizing a single material to detect multiple highly toxic gases can enhance safety across both industrial and non-industrial settings. Alternative HCN detection techniques—such as fluorometric, chromatographic, electrochemical, and luminescent methods—tend to be costly, time-intensive, and complex \cite{12,13,14,15}. Therefore, two-dimensional materials like MoS$_2$ offer a promising and efficient substitute. 

A computational study of pristine and doped MoS$_2$ for the sensing of HCN is outlined in two parts. The first part addresses the adsorption of HCN on the pristine MoS$_2$ monolayer and monolayers where one sulfur atom is replaced by a dopant X (X = P, N, Si, Al, B, Cl), resulting in P-, N-, Si-, Al-, B-, and Cl-doped MoS$_2$ monolayers, followed by a study of HCN interaction with these doped systems. Various properties, including structural, electrical, and optical characteristics, have been examined in detail. The evaluation of these properties led to the identification of suitable candidates for optical, electrochemical, and chemiresistive sensing applications. Optical sensors operate based on changes in various optical parameters, such as color, refractive index, absorption coefficient, dielectric function (including both real, $\epsilon_1$, and imaginary, $\epsilon_2$, components), and reflectance of the sensing material before and after HCN adsorption \cite{16}. Electrochemical sensors operate on the principle of charge transfer, i.e., the exchange of charge between the monolayer and the HCN molecule \cite{17}. Chemiresistive sensors operate based on the variations in electrical parameters of the sensing material, such as band gap, resistance, conductance or conductivity, work function, and electron emission current density, observed after the adsorption of the toxic HCN molecule \cite{18}.

A thorough investigation in the second part explains how increasing the doping by replacing more sulfur atoms in P-, Si-, and Al-doped MoS$_2$ monolayers can tune the adsorption energy, thus providing better control over the recovery time. A large recovery time, indicating a stronger and more stable interaction, can be shortened by increasing the temperature; however, this method has its own drawbacks. On the other hand, the doping strategy addresses these limitations and offers an additional advantage by enabling control over the adsorption strength on a specific material. A key finding, as explained in subsequent sections, is that increasing the number of dopants, i.e., by replacing more sulfur atoms, does not always enhance the adsorption strength.

\section{Computational Methodology}
Throughout this work, all density functional theory (DFT) calculations were perform
ed using the Spanish SIESTA package \cite{19,20}, which is considered a reliable tool for investigating two-dimensional systems. To treat the exchange–correlation interactions, the Generalized Gradient Approximation (GGA) in the form of Perdew–Burke–Ernzerhof (PBE) was employed \cite{21}. However, since the GGA-PBE functional is not well-suited for weakly bonded systems involving physisorption, we also employed the vdW-DF functional with the Vydrov–van Voorhis (VV10) kernel, specifically the vk-KBM functional, which is considered effective for calculating adsorption energies in systems dominated by van der Waals (vdW) interactions between the atoms \cite{22}. We used relativistic, norm-conserving Troullier–Martins pseudopotentials \cite{23} to describe the core electrons. The inclusion of relativistic effects inherently accounts for the spin component \cite{24}. The double-$\zeta$ plus polarization (DZP) basis set \cite{20} was employed, since it provides efficient performance with reasonable accuracy. An optimized $5 \times 5 \times 1$ Monkhorst–Pack k-point mesh was employed for Brillouin zone sampling \cite{25}. Despite being relatively coarse, this mesh yields reliable results owing to the use of a large supercell. A mesh cutoff energy \cite{20} of 300 Ry was adopted to ensure a well-converged real-space grid resolution. Structural relaxations were carried out using the conjugate gradient (CG) method \cite{20}, where atomic positions were optimized until the maximum residual force on each atom was below $0.01$ eV/\AA. The self-consistent field (SCF) calculations \cite{20} were considered converged when the total energy change fell below $10^{-5}$ eV.

\section{Results and Discussion}
\subsection{Optimised geometries of pristine and doped MoS\textsubscript{2} monolayers}
\begin{figure}[ht]
    \centering
    \includegraphics[width=0.9\textwidth]{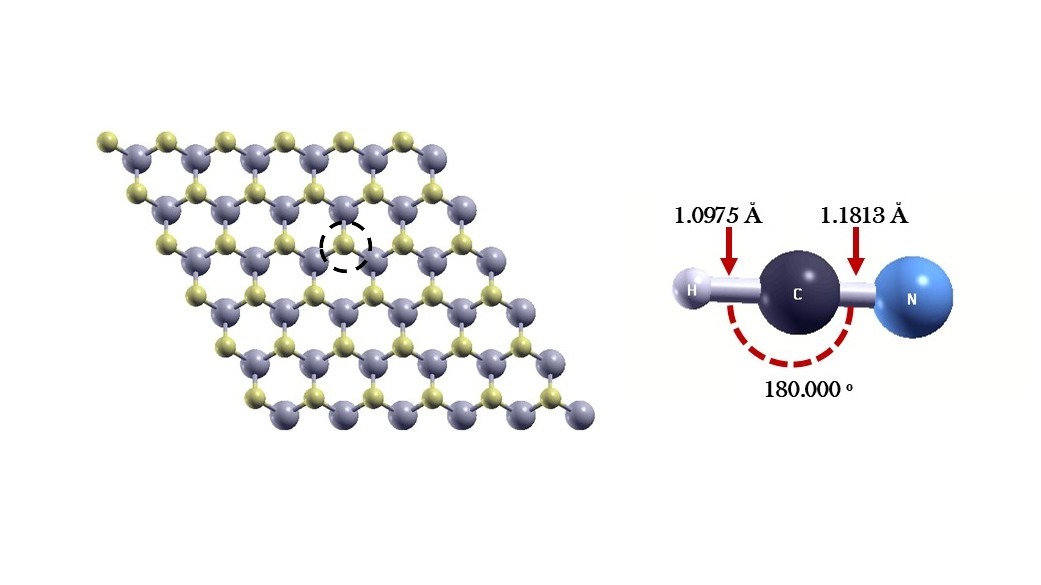}
    \caption{Optimized structure of the $6 \times 6 \times 1$ 1H-MoS$_2$ supercell (left) and HCN molecule (right). The circled sulfur atom is substituted with an X atom (X = P, N, Si, Al, B, Cl).}
    \label{fig:structure}
\end{figure}

Firstly, the structure of the MoS$_2$ monolayer (hexagonal phase, containing one Mo atom and two S atoms) was optimised. The optimized lattice constant was found to be 3.209\,\AA \cite{29}. The Mo--S bond length was 2.438\,\AA, and the S--Mo--S bond angle was approximately $81.055^\circ$. Subsequently, a vacuum of 20\,\AA{} was introduced along the \textit{z}-direction (to avoid interlayer interactions), and the system was extended to form a $6\times6\times1$ supercell containing 108 (36 Mo and 72 S) atoms. The supercell is sufficiently large to effectively eliminate mutual interactions between the HCN molecules. The resultant supercell structure is shown in Figure~\ref{fig:structure} (left). Also, the optimised structure of HCN molecule is shown in Figure~\ref{fig:structure} (right). 
To investigate the impact of doping, one sulfur atom (indicated with a black circle in Figure~\ref{fig:structure}, left) was substituted with an X atom, with the resultant bond lengths and band gaps detailed in Table~\ref{tab:bond_bandgap}. This particular doping method involves a dual process, initially creating a vacancy by the removal of the sulfur atom and subsequently substituting the dopant X atom (where X represents P, N, Si, Al, B, Cl) \cite{30}. Previous research \cite{30} has demonstrated that forming a sulfur vacancy within a pristine MoS$_2$ monolayer in two dimensions requires less energy than extracting a molybdenum atom. Experimentally, sulfur vacancies may be introduced using techniques such as chemical etching, hydrothermal synthesis, or electron beam irradiation \cite{30,31}.

Through bond-length calculations, it was determined that the Mo-X (where X represents P, N, Si, Al, B, Cl) bond lengths are shorter than the $Mo-S$ bond length, leading to the X-atoms being positioned below the MoS$_2$ plane along the $z$-axis while, for larger bond lengths, the X-atoms aligned above the plane. To ensure the accuracy of our findings, comparisons were made with values reported in the existing literature \cite{26,27,28,29,30,31,32,33,34,35,36,37,38}.

\begin{table}[H]
\centering
\caption{Bond length and band gap values for different MoS$_2$-based monolayers.}
\setlength{\tabcolsep}{4pt}
\renewcommand{\arraystretch}{1.15}
\begin{tabular}{|c|l|c|c|}
\hline
\textbf{Sr. No.} & \textbf{System} & \textbf{$d_{\mathrm{Mo\!-\!X}}$} & \textbf{$E_g$} \\
 &  & \textbf{(\AA)} & \textbf{(eV)} \\
\hline
1 & MoS$_2$     & 2.438 (2.415)\textsuperscript{\cite{30}} & 1.604 (1.68)\textsuperscript{\cite{37}} \\
2 & P--MoS$_2$  & 2.443 (2.43)\textsuperscript{\cite{28}} & 0.149 (0,1.512)\textsuperscript{\cite{28,36,41}} \\
3 & N--MoS$_2$  & 2.037 (2.08)\textsuperscript{\cite{27}} & 0.252 (0)\textsuperscript{\cite{38,41}} \\
4 & Si--MoS$_2$ & 2.392 (2.42)\textsuperscript{\cite{28}} & 0.823 (0.574, 0.72)\textsuperscript{\cite{28,41}} \\
5 & Al--MoS$_2$ & 2.580 (2.52)\textsuperscript{\cite{32}} & 0.312 (0.33)\textsuperscript{\cite{32}} \\
6 & B--MoS$_2$  & 2.135 (2.27)\textsuperscript{\cite{33}} & 0.226 (0, 1.51)\textsuperscript{\cite{33,41}} \\
7 & Cl--MoS$_2$ & 2.556 (2.51)\textsuperscript{\cite{33}} & 1.517 (1.68)\textsuperscript{\cite{35}} \\
\hline
\end{tabular}

\vspace{4pt}
\label{tab:bond_bandgap}
\end{table}


\subsection{Adsorption of HCN on pristine and doped monolayers}
\begin{figure}[ht]
    \centering
    \includegraphics[width=1.0\textwidth]{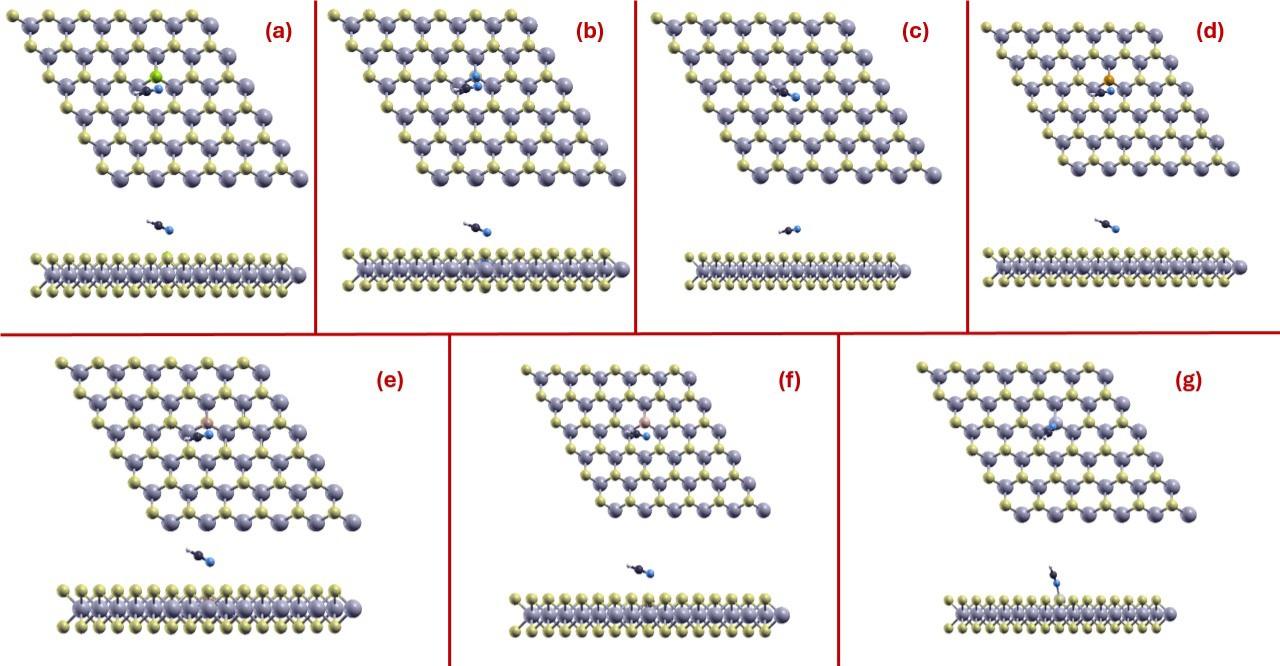}
    \caption{Optimised structure of $HCN$-molecule adsorbed on doped-monolayers of type (a) Cl-MoS$_2$, (b) N-MoS$_2$, (c) Pristine MoS$_2$, (d) P-MoS$_2$, (e) B-MoS$_2$, (f) Si-MoS$_2$, (g) Al-MoS$_2$ monolayer. The circled atom indicates the dopant atom in pristine monolayer.}
    \label{fig:adsorption}
\end{figure}

The adsorption of HCN molecules on both pristine and X-doped MoS$_2$ (where X represents P, N, Si, Al, B, Cl) monolayers has been examined. Initially, the HCN molecule was positioned approximately 2\,\AA{} above the monolayer surface, and structural relaxation was subsequently performed. Throughout the relaxation process, the HCN molecule adopted various orientations with respect to the monolayer. The final relaxed configuration of the HCN-adsorbed MoS$_2$ systems is depicted in Figure~\ref{fig:adsorption}, and the parameters relevant to the adsorption are presented in Table~\ref{tab:adsorption-data}.

\begin{table}[ht]
\centering
\caption{Calculated parameters for HCN adsorption on pristine and doped MoS\textsubscript{2} monolayers.}
\scriptsize
\renewcommand{\arraystretch}{1.2}
\begin{tabular}{|c|c|c|c|c|c|c|c|}
\hline
\textbf{Sr.} & \textbf{System} & \makecell{\textbf{$E_{\text{ads}}$ (GGA)} \\ \textbf{(eV)}} & \makecell{\textbf{$E_{\text{ads}}$ (vdW)} \\ \textbf{(eV)}} & \makecell{\textbf{Height} \\ \textbf{(\AA)}} & \makecell{\textbf{H--C} \\ \textbf{(\AA)}} & \makecell{\textbf{C$\equiv$N} \\ \textbf{(\AA)}} & \makecell{\textbf{Recovery} \\ \textbf{Time (s)}} \\
\hline
1 & Cl-MoS\textsubscript{2} & -0.0026 & -0.2219 & 2.62 & 1.0981 & 1.1805 & 5.34 $\times 10^{-9}$ \\
2 & N-MoS\textsubscript{2} & -0.046 & -0.2792 & 2.099 & 1.0973 & 1.1805 & 48.9 $\times 10^{-9}$ \\
3 & MoS\textsubscript{2} & -0.0357 & -0.233 & 2.84 & 1.0981 & 1.1808 & 6.78 $\times 10^{-9}$ \\
4 & P-MoS\textsubscript{2} & -0.077 & -0.2558 & 2.58 & 1.0981 & 1.1819 & 19.8 $\times 10^{-9}$ \\
5 & B-MoS\textsubscript{2} & -0.103 & -0.3052 & 2.58 & 1.0979 & 1.1804 & 134 $\times 10^{-9}$ \\
6 & Si-MoS\textsubscript{2} & -0.1998 & -0.4367 & 2.44 & 1.0994 & 1.1813 & 21.3 $\times 10^{-6}$ \\
7 & Al-MoS\textsubscript{2} & -1.399 & -1.6278 & 1.87 & 1.0995 & 1.1782 & 1.26 $\times 10^{15}$ \\

\hline
\end{tabular}
\label{tab:adsorption-data}
\end{table}

To study the strength of adsorption on a particular system, adsorption energy is calculated using the following relation:
\begin{equation}
    E_{\mathrm{ads}} = E_{\mathrm{complex}} - E_{\mathrm{monolayer}} - E_{\mathrm{HCN}}
\end{equation}
where \(E_{\mathrm{complex}}\), \(E_{\mathrm{monolayer}}\), and \(E_{\mathrm{HCN}}\) represent the total energies of the monolayer with the adsorbed HCN molecule, the isolated monolayer, and the isolated HCN molecule, respectively \cite{4,7,22,26,27}.
The adsorption energy (E\textsubscript{ads}) for pristine and doped-monolayers with GGA-PBE and vdW functional are reported in Table~\ref{tab:adsorption-data}. We have found that in both GGA-PBE and vdW functional, vdW functional gives more accurate values of the adsorption energies than former one \cite{22}. Negative adsorption energy corresponds to an exothermic process, while positive adsorption energy corresponds to an endothermic process \cite{26}.

It is observed that the adsorption energy (in magnitude) significantly more for doped monolayers as compared to pristine monolayer which results stronger interaction between the HCN-molecule and the doped-monolayers as compared to the pristine monolayer (with vdW functional). The enhanced value of adsorption energy (E\textsubscript{ads}) with vdW functional is due to the electronegativity of doped-atom. The electronegativities on the Pauling scale for S, P, N, Si, Al, B, and Cl are reported as 2.58, 2.19, 3.04, 1.90, 1.61, 2.04, and 3.16, respectively. Existing literature indicates that in MoS$_2$ in its pristine form, there is an equilibrium in charge transfer between Mo and S atoms, which ensures a uniform distribution of charge. The sulfur atoms effectively occupy the active sites available on the Mo atoms, which results in weak physisorption interactions with $HCN$ gas molecules\cite{26,27,30}. However, when a sulfur atom is substituted with a different dopant atom, this charge balance is disrupted, thereby exposing active sites on the Mo atoms and increasing the reactivity of the surface\cite{27}.

\begin{figure}[ht]
    \centering
    \includegraphics[width=0.8\linewidth]{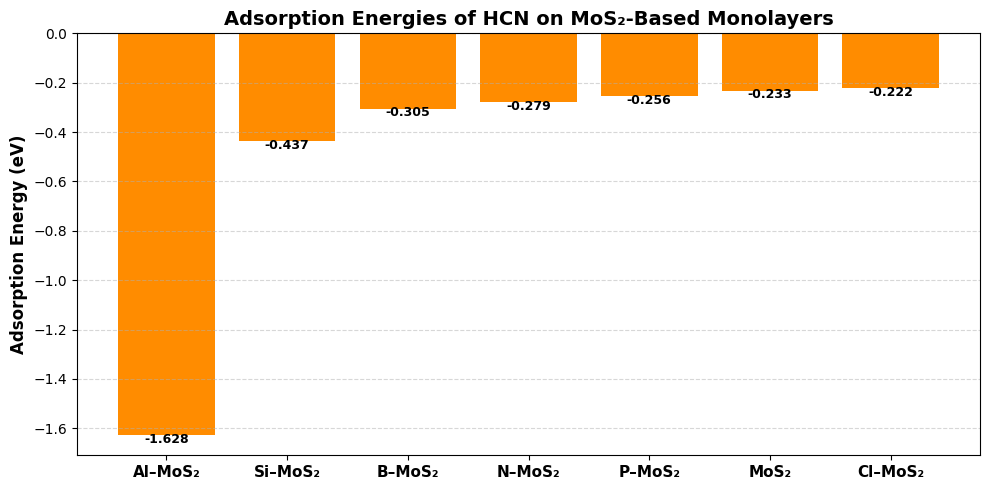}
    \caption{Adsorption energy ($E_{\text{ads}}$) trends for pristine and X-doped MoS$_2$ monolayers}
    \label{fig:Eads}
\end{figure}

Among six dopants, $Al$-atom has the lowest electronegativity $(1.61)$, which causes the greatest disturbance in the electronic environment\cite{28} and, consequently, the strongest interaction between the $HCN$-molecule and the $Al-MoS_2$ monolayer. As a result, the Al-doped MoS$_2$ monolayer exhibits the most negative (also most stable) adsorption energy, i.e., $-1.399$~eV (for GGA) and $-1.6278$~eV (for vdW). The large magnitude of the adsorption energy indicates chemisorption, as also illustrated in Figure~\ref{fig:adsorption}(e), where the $HCN$-molecule forms a bond with the Al-atom in Al-MoS$_2$. The adsorption strength of pristine and doped monolayers are reported in Table~\ref{tab:adsorption-data} and illustrated in Figure~\ref{fig:Eads} and the order of their energy is: Al-MoS$_2$ $>$ Si-MoS$_2$ $>$ B-MoS$_2$ $>$ N-MoS$_2$ $>$ P-MoS$_2$ $>$ MoS$_2$ $>$ Cl-MoS$_2$ (for vdW). This trend is consistent with some previously published works for different adsorbents \cite{27,28}. By comparing the results with the pristine MoS$_2$ monolayer, the adsorption strength of B-MoS$_2$, Si-MoS$_2$, and Al-MoS$_2$ monolayers is significantly enhanced, while that of the other systems remains almost unchanged.
The adsorption height and the bond lengths of the HCN molecule after the adsorption process are listed in Table~\ref{tab:adsorption-data}. The adsorption height is defined as the vertical distance between the topmost atom of the monolayer and the bottom-most atom of the HCN molecule. For the pristine and P-, N-, Si-, B-, and Cl-doped MoS$_2$ systems, the adsorption heights are greater than 2~\AA{}, indicating a weak \textbf{physisorption process}. In contrast, for the Al-doped MoS$_2$ system, the adsorption height of 1.87~\AA{} is smaller than the sum of the covalent radii of Al and N atoms (with the N atom oriented toward the monolayer), suggesting the formation of a chemical bond in this case, which further confirms the \textbf{chemisorption} nature of adsorption. The bond lengths of the adsorbed HCN molecule show only minor variations compared to the isolated state, further confirming the physisorption nature of the interaction.

Recovery time ($\tau$), which is an important parameter for evaluating the reusability and response efficiency of a gas sensor, refers to the time taken by sensor to reconfigure once the adsorbed molecules are released\cite{22}. The recovery time ($\tau$) is calculated using conventional transition state theory and is given by the following expression:
\begin{equation}
    \tau = \tau_0 \exp\left(\frac{E_{\text{ads}}}{k_B T}\right)
\end{equation}
where $\tau_0$ is the attempt frequency (typically $\sim 10^{-12}$~s) in the absence of radiation, $E_{\text{ads}}$ is the adsorption energy, $k_B$ is the Boltzmann constant, and $T$ is the temperature at which the HCN molecule desorbs from the monolayer (300~K in the present case)~\cite{22}. Desorption is the inverse process of adsorption and hence E\textsubscript{ads} denotes the energy barrier associated with desorption. It is evident from the equation that a larger magnitude of adsorption energy leads to a longer recovery time, while higher temperatures reduce the value of $\tau$ due to enhanced desorption rates. In our analyses, $\tau$ was calculated at ambient conditions (300~K). As illustrated in Table~\ref{tab:adsorption-data}, the recovery times for both pristine and doped monolayers (P-, N-, B-, and Cl-doped MoS$_2$) are measured in nanoseconds. However, the Si-doped monolayer exhibits a recovery time within the microsecond range (21.3~$\mu$s). In contrast, the Al-doped MoS$_2$ shows a significantly prolonged recovery period, spanning years, due to the substantial adsorption energy ($-1.6278$~eV), suggesting irreversible chemisorption.

From litrature, the recovery time was of the order of 175~s reported by Yang \textit{et al.}~\cite{39} for HCN adsorption on CuO nanostructures. In another work, presented by Singh \textit{et al.}~\cite{8}, who studied HCN adsorption on PANI/MMT and PANI/MMT–rGO composites materials, the sensors recovered automatically
within 21~s and 25~s, respectively. It is evident that the recovery time for both pristine and doped MoS$_2$ monolayers in our study are several orders of magnitude shorter than those found in experiments. Measuring the change in sensing characteristics within a single adsorption-desorption cycle might not be feasible for recovery durations in the millisecond range. The Si–MoS$_2$ monolayer is still appropriate for ultrafast response applications, though, if the sensing signal—which could result from charge transfer, a modification in the work function, or a change in the electronic structure—is sufficiently potent to be detected before desorption is finished. On the other hand, Al–MoS$_2$ cannot be efficiently reused because to its very long recovery period, making it unsuitable for sensing applications.

\subsection{Electronic Properties}
Here, we deals with the investigation of electronic properties of pristine and doped X-MoS$_2$ monolayers. The various parameters associated with the properties including variations in band gap, electrical conductance, or resistance after adsorption are tabulated in Table~\ref{tab:electronic-surface-properties}.
The band gap of the MoS$_2$ monolayer is consistent with the literature, as shown in Table~\ref{tab:bond_bandgap}. Some investigations define P-MoS$_2$ as a zero-band-gap material, while other studies report significant band gaps of around 1.51~eV and 1.71~eV (Table~\ref{tab:bond_bandgap}). Despite numerous papers presenting the band structures of N-MoS$_2$ monolayers, only Refs.~\cite{38,41} have reported N-MoS$_2$ as a zero-band-gap material. Similarly, for B-MoS$_2$, some studies have shown it to be a zero-band-gap material, while others have predicted a band gap as large as 1.51~eV. For the other monolayers, the band gap matches closely with previously reported values, as clearly shown in Table~\ref{tab:bond_bandgap}. The credibility of the generated monolayer structures is reinforced by the close agreement between our calculated bond lengths and band gaps and the existing data, notwithstanding these differences.

\begin{figure}[ht]
    \centering
    \includegraphics[width=1.0\textwidth]{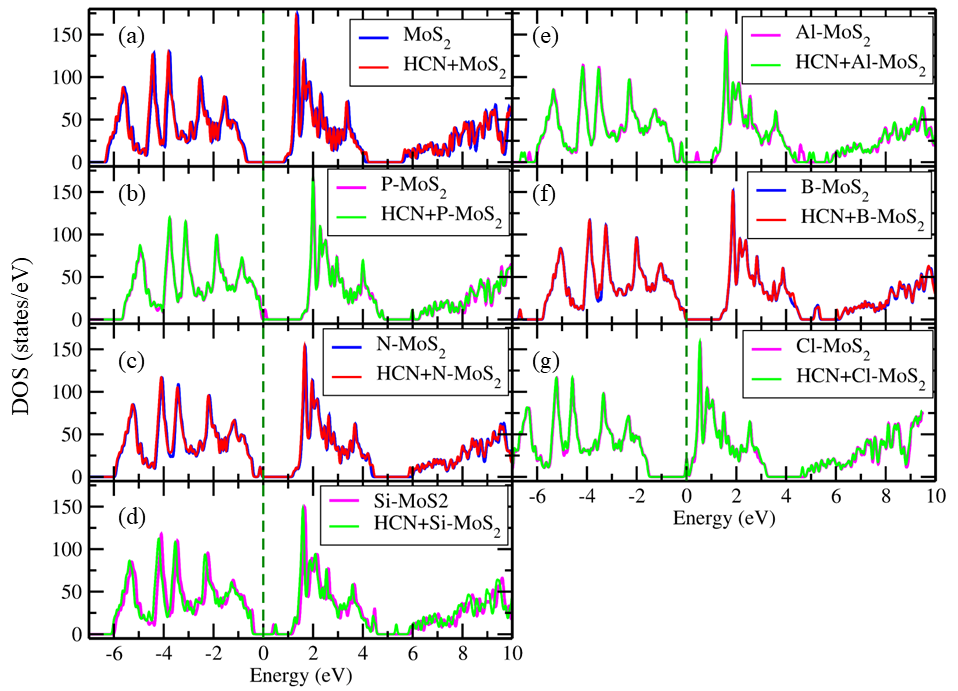}
\caption{Total density of states (TDOS) of (a) Pristine MoS$_2$, (b) P–MoS$_2$, (c) N–MoS$_2$, (d) Si–MoS$_2$, (e) Al–MoS$_2$, (f) B–MoS$_2$, and (g) Cl–MoS$_2$ monolayers before and after HCN adsorption. In all cases, the Fermi level is set to 0~eV and is represented by a vertical dotted green line.}
    \label{fig:dos}
\end{figure}

\begin{table}[ht]
\centering
\caption{Calculated values of HCN adsorption (before and after) on pristine and doped MoS\textsubscript{2} monolayers. $E_{g}^{\text{before}}$ and $E_{g}^{\text{after}}$ represents the band gap before and after adsorption of HCN, respectively; $\%\Delta E_g$ is the percentage change in band gap; $f_\sigma$ is the ratio of electrical conducitivity of doped monolayer after to before adsorption; $\Delta Q$ denotes the charge transfer (in $e$); $\phi_1$ and $\phi_2$ are the work functions before and after adsorption of HCN, respectively; $\Delta \phi$ is their difference; and $f_j$ represents the ratio of emission current density of ML after to before HCN adsorption.}
\scriptsize
\renewcommand{\arraystretch}{1.2}
\begin{tabular}{|c|c|c|c|c|c|c|c|c|c|c|}
\hline
\textbf{Sr.} & \textbf{System} & \makecell{$E_{g}^{\text{before}}$ \\ (eV)} & \makecell{$E_{g}^{\text{after}}$ \\ (eV)} & \makecell{\%$\Delta E_g$} & \textbf{$f_\sigma$} & \makecell{$\Delta Q$ \\ (e)} & \makecell{$\phi_1$ \\ (eV)} & \makecell{$\phi_2$ \\ (eV)} & \makecell{$\Delta \phi$ \\ (eV)} & \textbf{$f_j$} \\
\hline
1 & Cl–MoS\textsubscript{2} & 1.517 & 1.526 & 0.59 & 0.840 & 0.022 & 4.4988 & 4.4491 & 0.05 & 6.838 \\
2 & N–MoS\textsubscript{2} & 0.252 & 0.2596 & 3.02 & 0.863 & 0.030 & 5.6665 & 5.6087 & 0.06 & 9.355 \\
3 & MoS\textsubscript{2} & 1.604 & 1.6002 & 0.24 & 1.076 & -0.012 & 5.3414 & 5.3036 & 0.04 & 4.316 \\
4 & P–MoS\textsubscript{2} & 0.149 & 0.254 & 70.47 & 0.131 & 0.048 & 5.9628 & 5.8992 & 0.06 & 11.708 \\
5 & B–MoS\textsubscript{2} & 0.226 & 0.2284 & 1.29 & 0.955 & 0.024 & 5.8290 & 5.7849 & 0.04 & 5.507 \\
6 & Si–MoS\textsubscript{2} & 0.823 & 0.974 & 18.35 & 0.054 & 0.056 & 5.5963 & 5.4631 & 0.13 & 172.881 \\
7 & Al–MoS\textsubscript{2} & 0.312 & 0.2322 & 25.55 & 4.681 & 0.360 & 5.5110 & 5.3730 & 0.14 & 208.155 \\

\hline
\end{tabular}
\label{tab:electronic-surface-properties}
\end{table}
\begin{figure}[ht]
    \centering
    \includegraphics[width=1.0\textwidth]{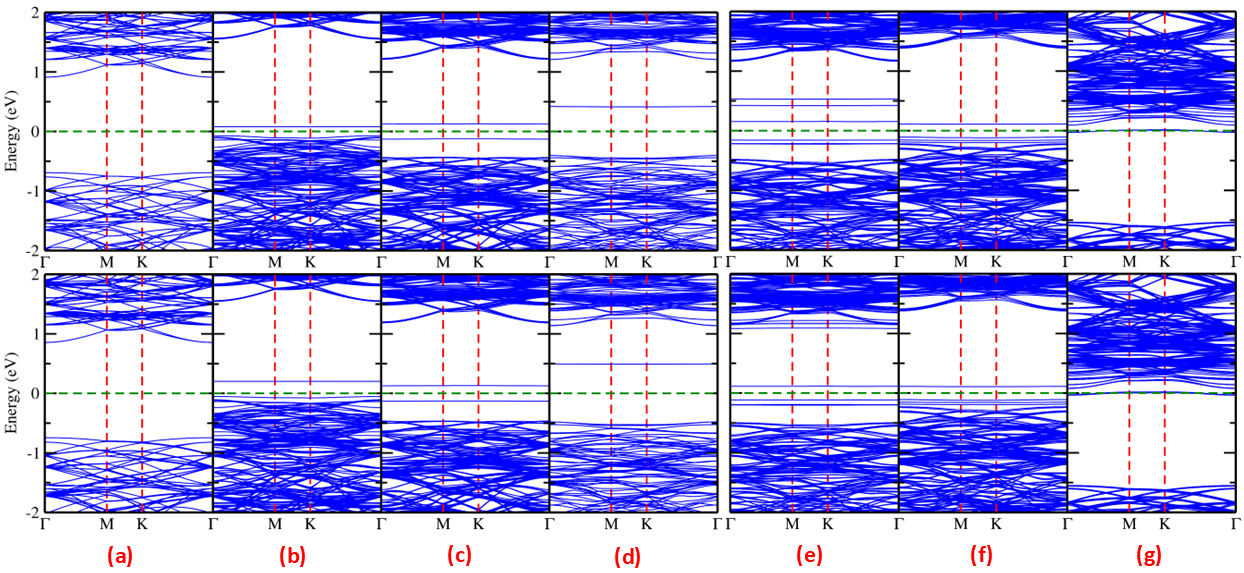}
\caption{Electronic band structures of (a) Pristine MoS$_2$, (b) P–MoS$_2$, (c) N–MoS$_2$, (d) Si–MoS$_2$, (e) Al–MoS$_2$, (f) B–MoS$_2$, and (g) Cl–MoS$_2$ monolayers before HCN adsorption (top) and after HCN adsorption (bottom). In all cases, the Fermi level is set to 0~eV and is represented by horizontal dotted green line. Vertical dotted red lines indicate the high-symmetry points in the Brillouin zone.}
    \label{fig:bands}
\end{figure}
The band structures of various systems before (top) and after (bottom) HCN adsorption are shown in Figure~\ref{fig:bands}. The total density of states (DOS) for pristine and doped-monolayers with the adsorption of $HCN$ are shown in Figure~\ref{fig:dos}. we observed that there is no significant change occured, except for a slight variation in the band gap, which supports a physisorption mechanism\cite{22,26}. However, for the $Al-MoS_2$ monolayer, a noticeable change in the DOS near the Fermi level is evident. The disappearance of states following HCN adsorption suggests a strong molecule substrate interaction, indicative of chemisorption, in contrast to the other systems where no substantial DOS modifications occur.

We observed a notable finding that the band gap decreases after adsorption for pristine monolayer of MoS$_2$ and Al-doped MoS$_2$, a minimal change in band gap occurs at the third decimal place, whereas it increases for all the other systems after adsorption of $HCN$. The reduction in band gap of $Al-MoS_2$ monolayer is likely due to strong adsorption, which aligns with the corresponding charge transfer and changes in the density of states. In contrast, the slight band gap widening in the other systems further supports a physisorption-dominated adsorption mechanism.

From the perspective of gas sensing applications, the percentage change in band gap ($\%\Delta E_g$) is an important parameter, which is calculated using the relation:
\begin{equation}
    \%\Delta E_g = \left( \frac{E_g^{\text{after}} - E_g^{\text{before}}}{E_g^{\text{before}}} \right) \times 100,
\end{equation}
where $E_g^{\text{before}}$ and $E_g^{\text{after}}$ represent the band gap values of the monolayer before (Table~\ref{tab:bond_bandgap}) and after (Table~\ref{tab:electronic-surface-properties}) adsorption of the HCN molecule, respectively~\cite{22}. The calculated values have been tabulated in Table~\ref{tab:electronic-surface-properties}. Notably, P–MoS$_2$, Si–MoS$_2$, and Al–MoS$_2$ show band gap changes of 70.47\%, 18.35\%, and 25.55\%, respectively. Other monolayers show very small percetange change in the bandgap.

From an experimental point of view, one can easily measure the electrical conductivity ($\sigma$) of a material, and it is well known that the conductivity of a semiconductor varies with its band gap according to the following relation:
\begin{equation}
    \sigma = \sigma_0 \exp\left(\frac{-E_g}{2k_B T}\right)
\end{equation}
Since there is a significant change in the band gap after the adsorption of an HCN molecule, the conductivity of the systems is also expected to vary accordingly. We have calculated the ratio of electrical conductivity of the monolayer after HCN adsorption to that before adsorption, denoted as $f_{\sigma}$ in Table~\ref{tab:electronic-surface-properties}. The calculated values show that ($f_{\sigma}$) have large value for pristine and Al-doped MoS$_2$ monolayers, while others has low values. This behavior can be attributed to the fact that the band gap decreases after adsorption for pristine and Al-doped MoS$_2$, whereas it increases for the other systems.

\begin{figure}[ht]
    \centering
    \includegraphics[width=0.9\textwidth]{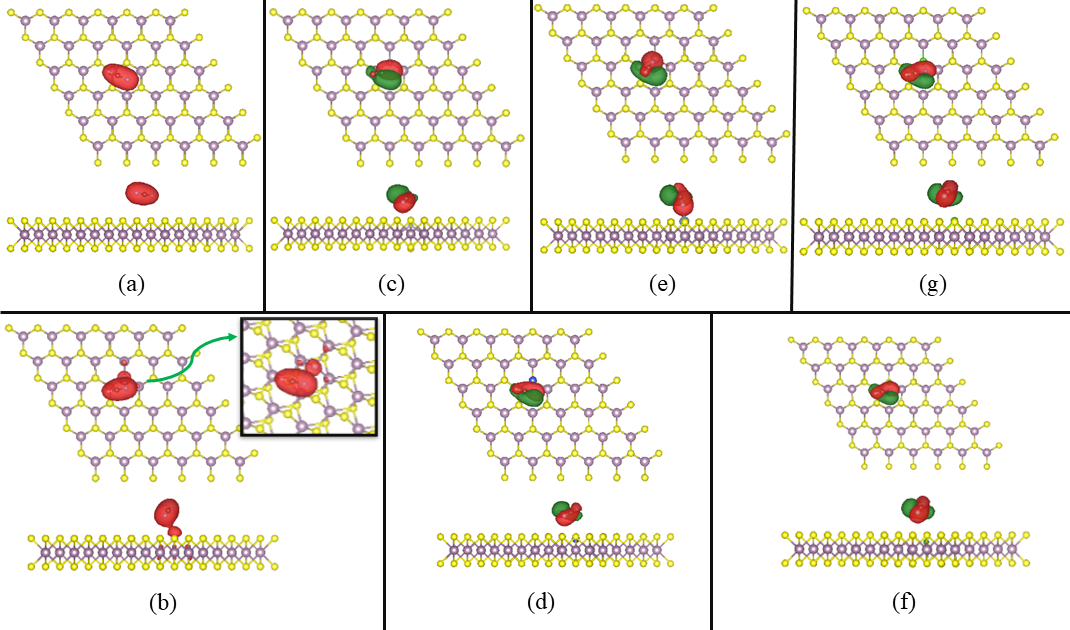}
    \caption{Charge density difference (CDD) plots illustrating the distribution of charge upon HCN adsorption on (a) pristine MoS$_2$, (b) P–MoS$_2$, (c) N–MoS$_2$, (d) Si–MoS$_2$, (e) Al–MoS$_2$, (f) B–MoS$_2$, and (g) Cl–MoS$_2$ monolayers. Red and green regions indicate charge accumulation and depletion, respectively, at an isosurface value of 0.00648~$e/\text{\AA}^3$.}
    \label{fig:cdd}
\end{figure}

Charge transfer is also an important parameter for assessing the interaction strength between the adsorbate and the adsorbent\cite{27,28,30}. The quantitative analysis of charge transfer was performed using Mulliken charge population, with the results summarized in Table~\ref{tab:electronic-surface-properties}. To calculate the charge transfer, we have used the relation:
\begin{equation}
    \Delta Q = Q\textsubscript{HCN}\textsuperscript{iso} - Q\textsubscript{HCN}\textsuperscript{ads}
\end{equation}
where $Q\textsubscript{HCN}\textsuperscript{iso}$ and $Q\textsubscript{HCN}\textsuperscript{ads}$ represent the total charge on the HCN molecule when isolated and when adsorbed on the respective monolayer. A positive $\Delta Q$ means that charge is transferred from the HCN molecule to the monolayer, whereas a negative $\Delta Q$ means that charge has been transferred from the monolayer to the HCN molecule \cite{2}. For a qualitative understanding and to better visualize regions of charge accumulation and depletion, charge density difference plots are provided in Figure~\ref{fig:cdd}.
\begin{figure}[ht]
    \centering
    \includegraphics[width=0.5\textwidth]{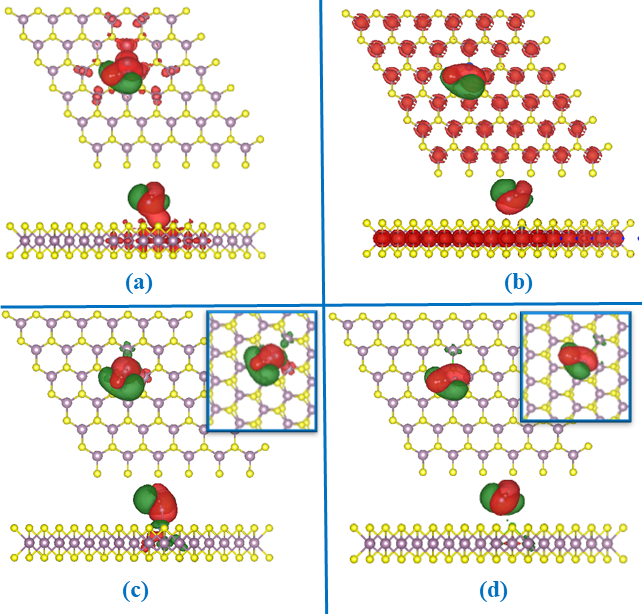}
    \caption{Charge density difference (CDD) plots illustrated in Figure~\ref{fig:cdd} with lower isosurface value of 0.001~$e/\text{\AA}^3$ for (a) P–MoS$_2$, (b) Si–MoS$_2$, (c) Al–MoS$_2$, and (d) Cl–MoS$_2$ monolayers}.
    \label{fig:cdd2}
\end{figure}
The charge density difference (CDD) is computed using the following relation:
\begin{equation}
    \Delta \rho = \rho\textsubscript{complex} - \rho\textsubscript{monolayer} - \rho\textsubscript{HCN}
\end{equation}
where $\rho_{\text{complex}}$, $\rho_{\text{monolayer}}$, and $\rho_{\text{HCN}}$ represent the charge density distributions of the combined HCN-monolayer system, the isolated monolayer, and the isolated HCN molecule, respectively \cite{28}. As reported in Table~\ref{tab:electronic-surface-properties}, in the pristine monolayer, charge is transferred from the monolayer to the HCN molecule, which is also evident from the CDD plots. Doped systems also show small charge transfer; however, their effects on the monolayer are not clearly visible in Figure~\ref{fig:cdd}, as the isosurface value of 0.00648~$e/\text{\AA}^3$ is relatively high. Therefore, we generated all the CDD plots with a lower isosurface value of 0.001~$e/\text{\AA}^3$. At this value, only P-doped, Si-doped, Al-doped, and Cl-doped monolayers showed noticeable charge variation in the monolayer, as shown in Figure~\ref{fig:cdd2}. Other systems, having even smaller values of charge transfer, are therefore not shown.

Almost every system, other than pristine MoS$_2$, shows some charge accumulation in the monolayer, further confirming the quantitative data given in Table~\ref{tab:electronic-surface-properties}. A key observation is that the charge transfer values are consistent with the corresponding adsorption energy values. For instance, in the case of the Al–MoS$_2$ monolayer, which exhibits the strongest adsorption energy, a charge transfer of $0.36~e$ to the monolayer is observed—the highest among all the systems studied. This significant charge transfer further supports the conclusion that the adsorption process in this system is chemisorption, in contrast to the other systems where physisorption is predominant. 

In Si-doped MoS$_2$, charge has been accumulated on almost every Mo atom in the supercell. In the Al-doped monolayer, charge accumulation is observed on the aluminium atom, while the neighboring three Mo atoms show a decrease in charge density, indicating bond formation and chemisorption. In the Cl-doped monolayer, a similar behavior is seen, where a noticeable charge is accumulated on the Cl atom and a slight decrease in charge density is observed near the Mo atom, supporting the theoretical explanation based on electronegativity differences\cite{28}. However, due to the small magnitude of charge transfer, the overall effect is subtle. In the P-doped monolayer, the CDD plots clearly show charge accumulation on the neighboring atoms of the HCN molecule in the monolayer.

The adsorption of HCN induces noticeable charge transfer, which subsequently alters the work function of the monolayers. The variation in the work function is defined as
\begin{equation}
   \Delta \phi = \phi_1 - \phi_2,
\end{equation}
where $\phi_1$ and $\phi_2$ denote the work function of the isolated monolayer and of the monolayer with the HCN molecule adsorbed on its surface, respectively. The work function, $\phi$, is the minimum energy required to remove an electron from the Fermi level of the material to a point just outside the surface (the vacuum level). It is calculated as
\begin{equation}
    \phi = V_{\mathrm{vac}} - E_F,
\end{equation}
where $V_{\mathrm{vac}}$ is the electrostatic potential in the vacuum region far from the surface, and $E_F$ is the Fermi energy of the system. For simplicity, the vacuum level is often taken as the reference zero. A larger value of $\phi$ indicates a higher energy barrier for electron emission from the surface \cite{22}.

The work functions before adsorption ($\phi_1$), after adsorption ($\phi_2$), and the change in work function ($\Delta\phi$) are presented in Table~\ref{tab:electronic-surface-properties}. The magnitude of the change in work function serves as an indicator of a material's sensitivity in gas-sensing applications. Following HCN adsorption, a reduction in the work function is observed for all monolayers compared to their pristine forms. Notably, the Si--MoS$_2$ and Al--MoS$_2$ monolayers exhibit the largest changes of 0.13~eV and 0.14~eV, respectively. These values lie within the experimentally detectable range, highlighting the potential of these two systems for sensor applications based on work-function modulation.

Using the value of the work function, one can also calculate the electron emission current density ($j$). This can be done using the Richardson equation, which relates $j$ to the work function ($\phi$) as follows:
\begin{equation}
    j = A T^2 \exp\left(-\frac{e\phi}{k_{\text{B}} T}\right)
\end{equation}
where \(A\) is the Richardson constant and its value is \(1.2\times 10^{6}\,\mathrm{A/(m}^{\mathrm{2}}\mathrm{\cdot K}^{\mathrm{2}}\mathrm{)}\), \(T\) is the absolute temperature in Kelvin, \(\phi \) is the material's work function in electron-volts (eV), representing the minimum energy required to remove an electron from the surface to infinity. \(e\) is the elementary charge (\(e\approx 1.602\times 10^{-19}\,\mathrm{C}\)), \(k_{\text{B}}\) is the Boltzmann constant (\(k_{\text{B}}\approx 1.381\times 10^{-23}\,\mathrm{J/K}\)). 

The work function following HCN adsorption leads to an increase in the electron emission current density ($j$) compared with pristine monolayer. The ratio of $j$ after adsorption to that before adsorption is calculated and tabulated in Table~\ref{tab:electronic-surface-properties} as $f_j$. Among the studied systems, $Al-MoS_2$ shows the highest enhancement in $j$ after HCN adsorption, exhibiting an increase of approximately 208 times compared to its pristine form. Similarly, $Si-MoS_2$ and $P-MoS_2$ show about 172-times and 11-times increase, respectively.

\subsection{Optical properties}
Time-dependent perturbations of the ground-state electronic states have been used to describe how photons interact with electrons. Transitions between the occupied and unoccupied electronic states are caused by the electric field connected to the incident photons. However, our main focus is on how the optical characteristics of the monolayers change upon the adsorption of the HCN molecule for gas-sensing applications.
Here, we have calculated the absorption coefficient, refractive index, and the real and imaginary parts of the dielectric function. Among the seven systems considered, the P–MoS$_2$ and Si–MoS$_2$ monolayers exhibited the most favorable recovery times; therefore, our discussion is limited to these systems along with the Al–MoS$_2$ monolayer. The Al–MoS$_2$ system is included even though it shows chemisorption and a very large recovery time, as it provides useful insight into the effect of doping on adsorption. 

\begin{figure}[ht]
    \centering
    \includegraphics[width=0.75\textwidth]{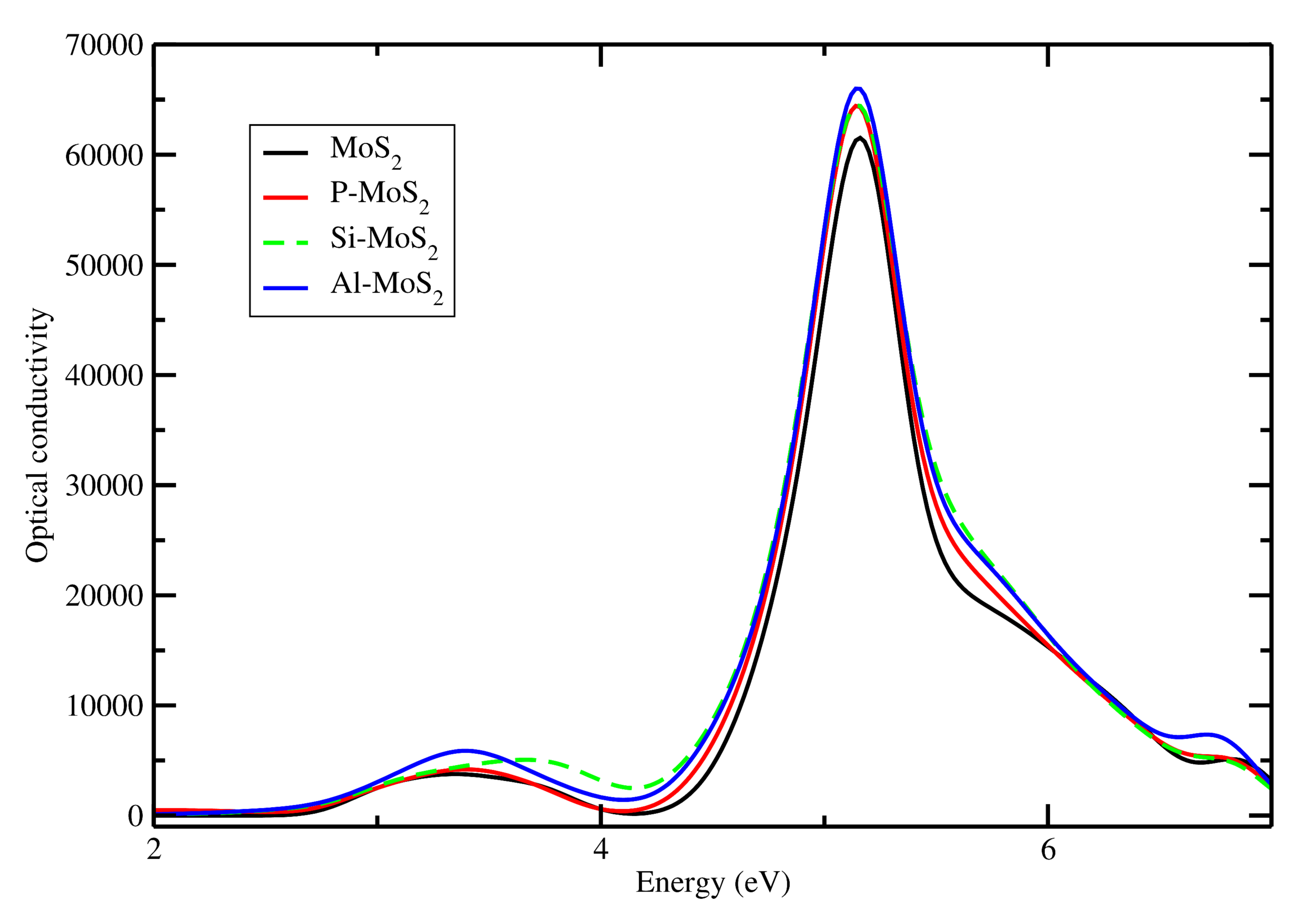}
    \caption{Calculated optical conductivity of pristine MoS$_2$ (black), P-MoS$_2$ (red), Si-MoS$_2$ (dotted green) and Al-MoS$_2$ (blue) monolayers.}
    \label{fig:optical_conductivity}
\end{figure}

The optical conductivity of a material represents the increase in the number of charge carriers (electrons and holes) when light of a certain frequency is incident on it. The optical conductivity of the pristine MoS$_2$ monolayer is found to increase after doping with X atoms, as shown in Figure~\ref{fig:optical_conductivity}. Among all the doped systems, $Al-MoS_2$ exhibits the highest value of optical conductivity. The peak value of optical conductivity occured at photon energy around $5~eV$ for all four systems and it is observed most for $Al-MoS_2$ and least for pristine monolayer.

\begin{figure}[ht]
    \centering
    \includegraphics[width=1.0\textwidth]{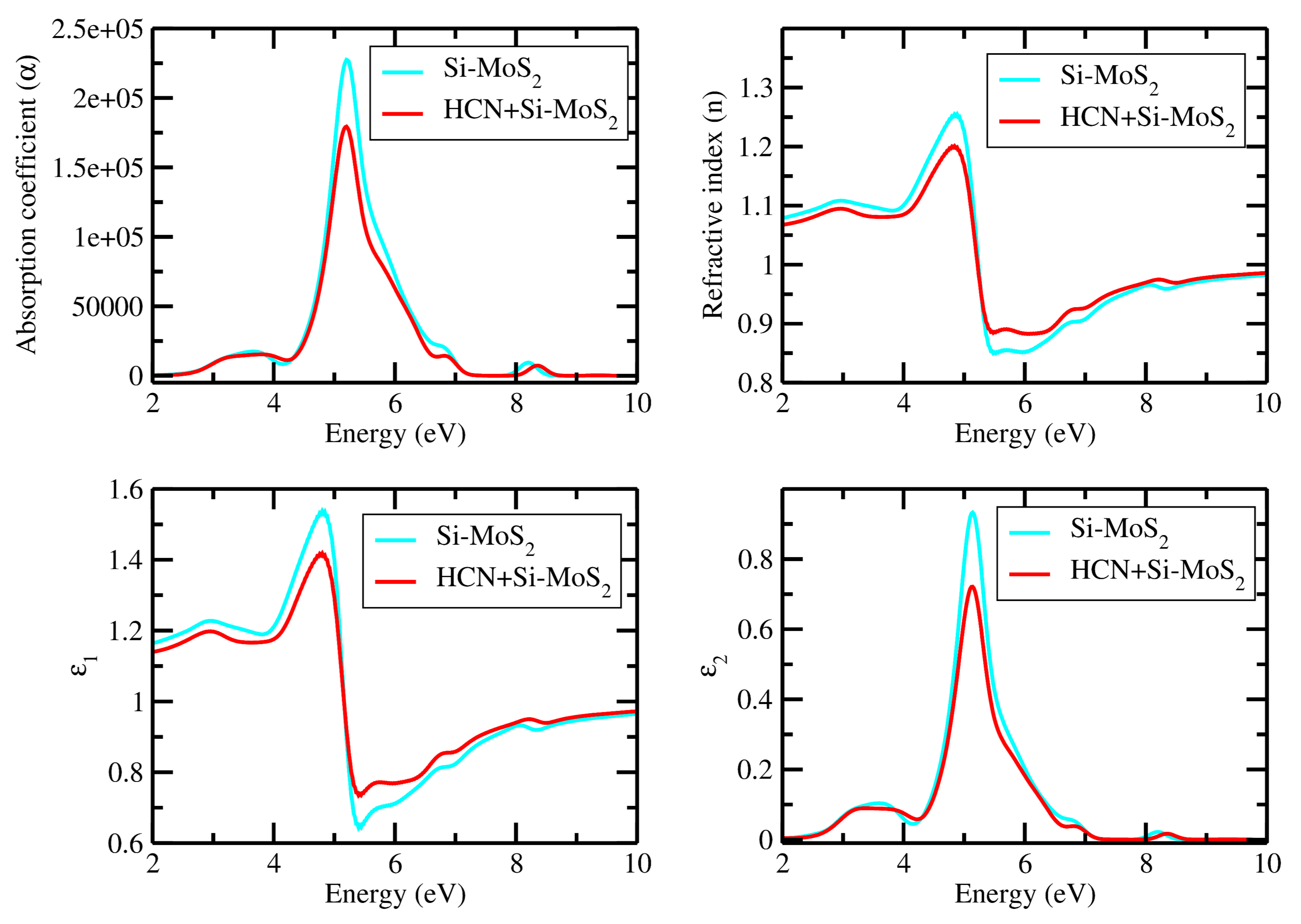}
\caption{Optical properties of pristine Si–MoS$_2$ monolayer (turquoise) and HCN-adsorbed Si–MoS$_2$ monolayer (red).}
    \label{fig:optical_si}
\end{figure}

\begin{table}[ht]
\centering
\caption{Percentage change in optical parameters of different monolayers after HCN adsorption.}
\label{tab:optical-properties}
\renewcommand{\arraystretch}{1.2}
\begin{tabular}{cccccc}
\hline
\textbf{Sr. No.} & \textbf{System} & \textbf{\%$\Delta\alpha$} & \textbf{\%$\Delta n$} & \textbf{\%$\Delta\varepsilon_1$} & \textbf{\%$\Delta\varepsilon_2$} \\ 
\hline
1 & MoS$_2$     & 27.04                  &3.59                   & 3.03                  &26.87                   \\ 
2 & P–MoS$_2$   &25.90\cite{26}                   &3.18                   & 7.20                  &25.93                   \\ 
3 & Si–MoS$_2$  &21.12                   &4.30                   & 7.78                  &22.68                   \\ 
4 & Al–MoS$_2$  &26.70                   &4.81                   &8.38                   &27.52                   \\ 
\hline
\end{tabular}
\end{table}

Next, we will see how the optical properties change when an HCN molecule is adsorbed on the monolayer. Figure~\ref{fig:optical_si} presents a comparison of the absorption coefficient ($\alpha$), refractive index ($n$), and the real ($\varepsilon_1$) and imaginary ($\varepsilon_2$) parts of the dielectric function of the Si–MoS$_2$ monolayer before and after HCN adsorption. The percentage change in the peak values of these optical properties was also evaluated. 

The absorption coefficient decreased by 19.02\%, the refractive index dropped by 6.61\%, the real part of the dielectric function ($\varepsilon_1$) decreased by 11.54\%, and the imaginary part ($\varepsilon_2$) showed a reduction of 22.68\%. These variations were observed within the photon energy range of 4.70–5.30~eV, corresponding to the ultraviolet (UV) region of the electromagnetic spectrum. Furthermore, after adsorption, the peaks of all optical properties shifted slightly towards lower energy values, indicating a red shift. Such changes are significant enough to be detected by UV-based optical sensors, suggesting the potential of Si–MoS$_2$ as an effective material for UV-responsive HCN sensing. A similar trend of decrease in optical properties was also observed for the other monolayers. The variations in the optical properties of pristine MoS$_2$, P–MoS$_2$, and Al–MoS$_2$ are provided in the Supporting Information (S1). The calculated values are summarized in Table~\ref{tab:optical-properties}. $\%\Delta\alpha$ represents the percentage change in the absorption coefficient, $\%\Delta n$ denotes the percentage change in the refractive index, while $\%\Delta\varepsilon_1$ and $\%\Delta\varepsilon_2$ correspond to the percentage changes in the real and imaginary parts of the dielectric function, respectively, after HCN adsorption relative to the pristine monolayer. 

\section{Recovery Time Modulation through Doping}
As discussed earlier, recovery time is a critical parameter for the application of gas sensors, as it ensures the reusability of the sensor. As per Eq.~(2), it can be controlled by temperature, and there is a lot of literature regarding this~\cite{22,28,30}. For instance, the recovery time for Al--MoS$_2$ can be reduced from around 40 million years at 300~K to 9 years at 400~K and further to 2.8 hours at 500~K. However, high temperatures have their own limitations, which include the stability of the adsorbing material and constraints in environments where very high temperatures are not desirable. Therefore, achieving recovery times within a suitable range at room temperature is essential. A. Sharma et al.~\cite{30}, in their work, have suggested the application of an electric field for desorbing the adsorbed phosgene molecule from a silicon-doped MoS$_2$ monolayer. However, there is one more parameter in equation~2 that one can play with — the adsorption energy, $E_{\text{ads}}$. In this section, it is discussed how we can control the adsorption energy by doping.

Based on our observations so far, P–MoS$_2$ and Si–MoS$_2$ monolayers have proved themselves as promising HCN gas-sensing materials. Therefore, this section will be limited to the study of these systems. The P–MoS$_2$ monolayer, apart from showing a large change in bandgap, significant reduction in electrical conductivity, noticeable change in electron emission current density, and significant changes in optical properties after HCN adsorption, has a recovery time in the nanosecond range. Therefore, this system has been chosen to investigate whether the $\tau$ value can be increased or not. The Si–MoS$_2$ monolayer has proved itself in almost every aspect, and its microsecond-scale $\tau$ is quite appropriate for applications where very fast detection is required. Along with these two systems, we have also chosen the Al–MoS$_2$ monolayer, as it has shown significant changes in all of its properties; however, it has a very large recovery time. The extremely high recovery time in Al–MoS$_2$ indicates a highly stable system after the adsorption of the HCN molecule, rendering it unsuitable for gas-sensing applications due to poor desorption characteristics. Hence, it will be interesting to see if we can bring down this recovery time.

\begin{figure}[ht]
    \centering
    \includegraphics[width=1.0\textwidth]{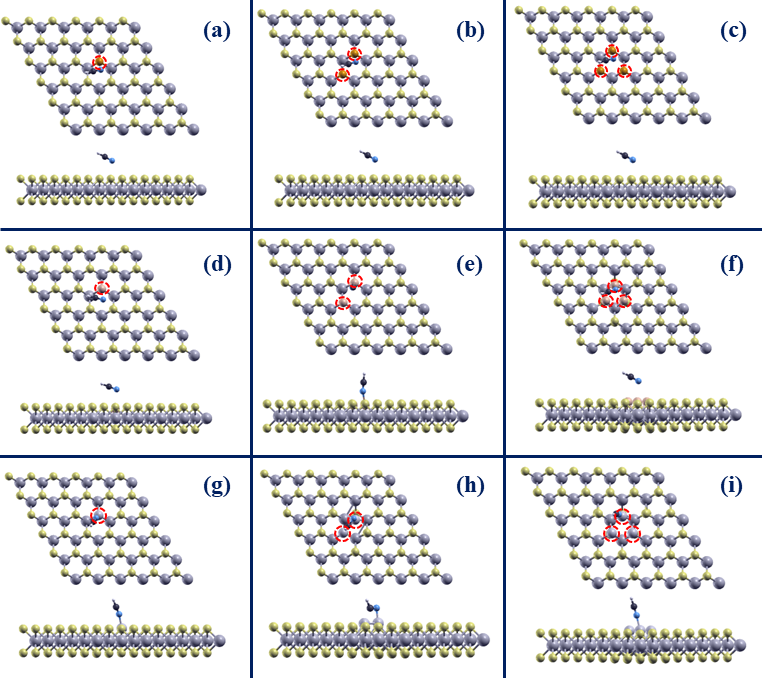}
\caption{Fully relaxed structures of HCN adsorbed on (a) P–MoS$_2$, (b) 2P–MoS$_2$, (c) 3P–MoS$_2$, (d) Si–MoS$_2$, (e) 2Si–MoS$_2$, (f) 3Si–MoS$_2$, (g) Al–MoS$_2$, (h) 2Al–MoS$_2$, and (i) 3Al–MoS$_2$ monolayers. The atoms marked by red dotted circles represent the corresponding dopant atoms.}
    \label{fig:2.1}
\end{figure}

At first, the doping concentration was increased in the P–MoS$_2$ monolayer, as this system has already been studied in detail for different adsorbates in previously published works~\cite{27,36}. Accordingly, one more S atom in the P–MoS$_2$ monolayer was substituted by a phosphorus atom, and the structure was relaxed. Then, the HCN molecule was adsorbed on the monolayer, and the corresponding properties were studied. Here, only those parameters that are important for recovery time and, therefore, for adsorption are discussed. One can refer to the supplementary information for detailed tables. The same procedure was followed when one more sulfur atom was replaced by a phosphorus atom, as shown in Figure~\ref{fig:2.1}(b,c). Similar calculations were carried out for Si–MoS$_2$ and Al–MoS$_2$ monolayers as well. The corresponding parameters are summarized in Table~\ref{tab:2}.

\begin{table}[htbp]
    \centering
    \caption{Variation of key adsorption parameters with increasing doping in MoS$_2$ monolayers.}
    \scriptsize
    \begin{tabular}{|c|c|c|c|c|c|c|c|}
        \hline
        \textbf{Sr.} & \textbf{System} & \makecell{$E_{\text{ads}}$ \\ (GGA) \\ (eV)} & \makecell{$E_{\text{ads}}$ \\ (vdW) \\ (eV)} & \makecell{Height \\ (\AA)} & \makecell{Recovery time, $\tau$ \\ (s)} & \makecell{H--C \\ (\AA)} & \makecell{C$\equiv$N \\ (\AA)} \\
        \hline
        1 & P–MoS$_2$   & -0.077  & -0.2558 & 2.58 & 19.8 $\times 10^{-9}$  & 1.0981 & 1.1819 \\
        2 & 2P–MoS$_2$  & -0.14   & -0.3570 & 2.56 & 1.004 $\times 10^{-6}$  & 1.0981 & 1.1807 \\
        3 & 3P–MoS$_2$  & -0.12   & -0.3457 & 2.57 & 0.641 $\times 10^{-6}$ & 1.0985 & 1.1820 \\
        4 & Si–MoS$_2$  & -0.1998 & -0.4367 & 2.44 & 21.3 $\times 10^{-6}$ & 1.0994 & 1.1813 \\
        5 & 2Si–MoS$_2$ & -0.906  & -1.08 & 1.62 & 1.38 $\times 10^{6}$ & 1.0980 & 1.1859   \\
        6 & 3Si–MoS$_2$ & 0.0132 & -0.143 & 2.74 & 2.52 $\times 10^{-10}$ & 1.0983 & 1.1804   \\
        7 & Al–MoS$_2$  & -1.399  & -1.6278 & 1.87 & 1.26 $\times 10^{15}$ & 1.0995 & 1.1782 \\
        8 & 2Al–MoS$_2$ & -1.577  & -1.6519 & 1.84 & 2.39 $\times 10^{15}$ & 1.1273 & 1.2996 \\
        9 & 3Al–MoS$_2$ & -0.635  & -0.7262 & 1.94 & 1.74 & 1.1003 & 1.1798 \\
        \hline
    \end{tabular}
\label{tab:2}
\end{table}

The adsorption energies for 2P–MoS$_2$, 2Si–MoS$_2$, and 2Al–MoS$_2$ monolayers increased as compared to P–, Si–, and Al–MoS$_2$ monolayers, respectively, suggesting a stronger interaction between the HCN molecule and the monolayer when two sulfur atoms are replaced by X (X = P, Al, Si) atoms. It can be seen from Figure~\ref{fig:2.1}(e) and (f) that the N atom of the HCN molecule has formed a bond with the Si and Al atoms of the 2Si– and 2Al–doped monolayers, respectively, indicating chemisorption. The HCN molecule undergoes significant structural changes upon adsorption on 2Al–MoS$_2$, as shown in Table~\ref{tab:2}. 

Next, three sulfur atoms were replaced in a symmetric manner, as shown in Figure~\ref{fig:2.1}(c), (d), and (i), leading to the formation of 3P–, 3Si–, and 3Al–MoS$_2$ monolayers. When adsorption was studied on these monolayers, it was found that the adsorption energy decreased in all three cases. The reduction in adsorption energy is not very significant for the 3P–MoS$_2$ monolayer; however, for 3Si– and 3Al–MoS$_2$, the adsorption energy decreased to values even lower than those of Si– and Al–MoS$_2$ monolayers, respectively. These systems now exhibit physisorption behavior. Even in the case of aluminum doping, which previously showed strong chemisorption, physisorption is now observed with an adsorption energy of $-0.7262$~eV. This can be clearly seen in Figure~\ref{fig:2.1}, as no bond formation takes place. Thus, symmetric doping of this type can serve as an effective approach to control the adsorption energy strength in these monolayers.

The adsorption energy values represent the strength of adsorption, which can also be verified from the height values given in Table~\ref{tab:2}. A stronger adsorption corresponds to a smaller adsorption distance, while weaker adsorption is associated with a larger adsorption height. It is important to note that this height is calculated as the vertical distance between the topmost atom of the monolayer under study and the bottommost atom of the HCN molecule.

\begin{figure}[htbp]
    \centering
    \includegraphics[width=1.0\linewidth]{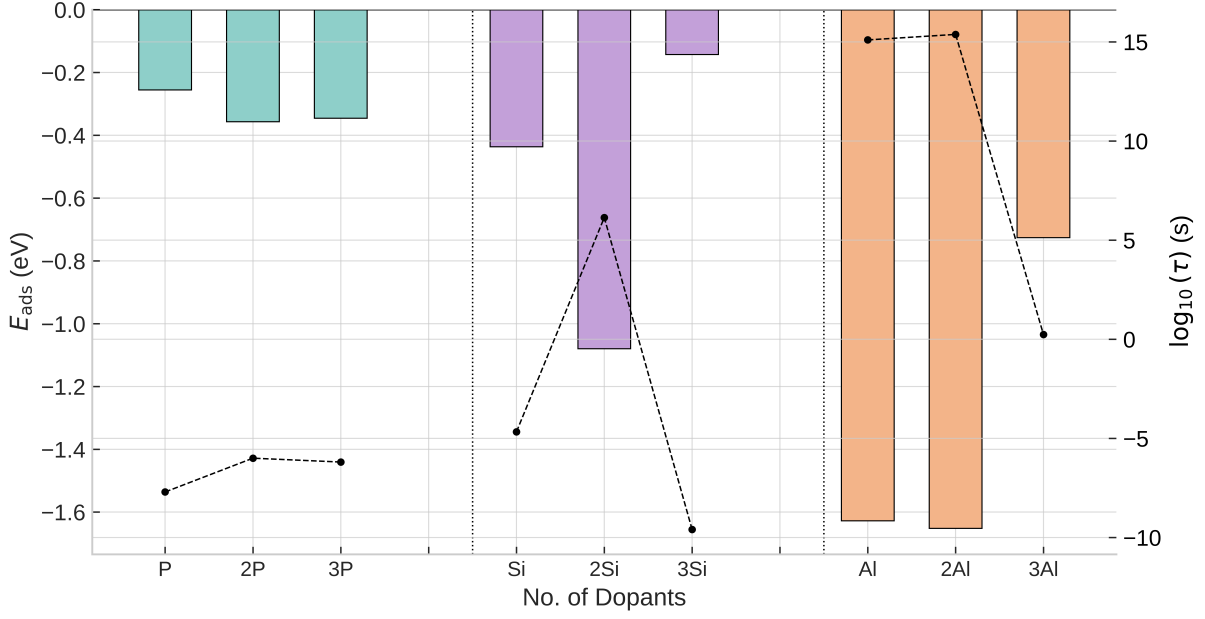}
    \caption{
        Variation of adsorption energy ($E_{\text{ads}}$) and recovery time ($\tau$) of HCN molecule on P-, Si-, and Al-doped MoS$_2$ monolayers with increasing number of dopants. 
        The bars represent adsorption energy values ($E_{\text{ads}}$ in eV), while the dashed line (right axis) shows the logarithm of recovery time [$\log_{10}(\tau)$ in seconds].
    }
    \label{fig:Eads_tau}
\end{figure}

The recovery time ($\tau$) for 2P–MoS$_2$ is calculated to be on the order of microseconds, while for 3P–MoS$_2$, it further decreases to a fraction of a microsecond. In the case of 2Si–MoS$_2$, the recovery time extends to several days, which is undesirable for practical gas-sensing applications. This long recovery time arises from the strong, nearly irreversible adsorption process. However, for 3Si–MoS$_2$, $\tau$ reduces drastically to the sub-nanosecond range. Thus, in silicon-doped systems, the variation in doping concentration alters the recovery time in a manner similar to phosphorus doping, although the resulting values are not favorable for sensing applications. For 2Al–MoS$_2$, the recovery time increases due to stronger adsorption. Interestingly, for 3Al–MoS$_2$, $\tau$ is reduced to approximately 1.74~seconds. Therefore, through appropriate and strategic doping, it is possible to effectively control the adsorption strength and, consequently, the recovery behavior of these monolayers. Figure~\ref{fig:Eads_tau} illustrates how adsorption strength and sensor response dynamics evolve as the dopant concentration increases for each dopant type.

\begin{table}[H]
\centering
\caption{Calculated electronic parameters for various X–MoS$_2$ (X = P, Si, Al) monolayers before and after HCN adsorption.}
\label{tab:electronic-parameters}
\renewcommand{\arraystretch}{1.2}
\begin{tabular}{c c c c c c c c c}
\hline
\textbf{Sr. No.} & \textbf{System} & \textbf{$E_g^{\text{before}}$} & \textbf{$E_g^{\text{after}}$} & \textbf{\%$\Delta E_g$} & \textbf{$\Delta Q$} & \textbf{$\phi_1$} & \textbf{$\phi_2$} & \textbf{$\Delta\phi$} \\
 &  & (eV) & (eV) &  & (e) & (eV) & (eV) & (eV) \\
\hline
1 & P–MoS$_2$   &0.149   &0.254   &70.47  &0.048   &5.9628  &5.8992  &0.06  \\
2 & 2P–MoS$_2$  &0.181\cite{36}   &0.267   &47.51  &0.047   &5.9377  &5.8456  &0.09  \\
3 & 3P–MoS$_2$  &0.2523  &0.3633  &43.99  &0.05    &5.9894  &5.9232  &0.07  \\
4 & Si–MoS$_2$  &0.823   &0.974   &18.35  &0.056   &5.5963  &5.4631  &0.13  \\
5 & 2Si–MoS$_2$ &0.5752  &0.503   &12.55  &0.298   &5.6335  &5.4691  &0.16  \\
6 & 3Si–MoS$_2$ &0.623   &0.60    &3.69   &0.044   &5.5160  &5.4287  &0.09  \\
7 & Al–MoS$_2$  &0.3119  &0.2322  &25.55  &0.36    &5.5110  &5.3730  &0.14  \\
8 & 2Al–MoS$_2$ &0.4398  &0.6352  &44.43  &-0.014  &5.3589  &5.2273  &0.132  \\
9 & 3Al–MoS$_2$ &0.1916  &0.1981  &3.39   &0.299   &5.2694  &5.0951  &0.174  \\
\hline
\end{tabular}
\end{table}

\begin{figure}[ht]
    \centering
    \includegraphics[width=0.7\textwidth]{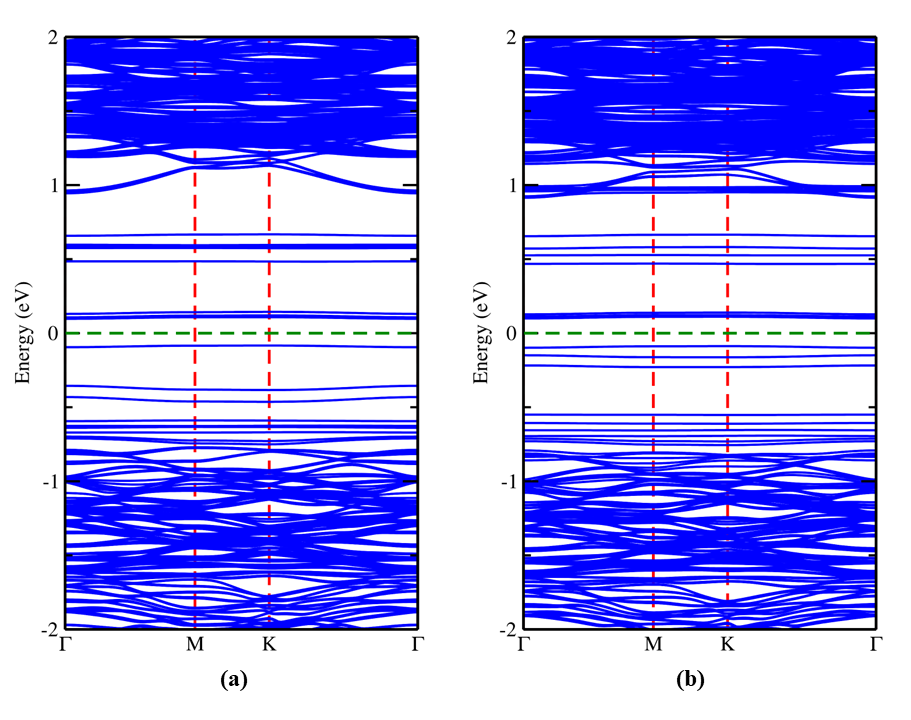}
\caption{Band structure of 3Al-MoS$_2$ before (a) and after (b) HCN adsorption. The Fermi level is shifted to 0~eV, indicated by the dotted green line.}
    \label{fig:2.2}
\end{figure}

\begin{figure}[H]
    \centering
    \includegraphics[width=1.0\textwidth]{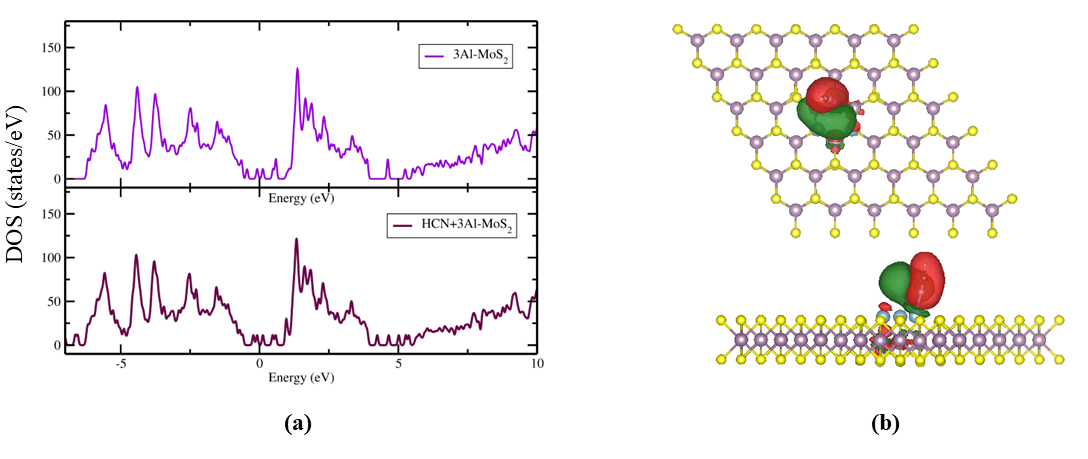}
    \caption{(a) Density of states (DOS) of the 3Al–MoS$_2$ monolayer before (top) and after (bottom) HCN adsorption. The Fermi level is aligned to 0~eV, marked by the dotted vertical green line. (b) Charge density difference (CDD) plot of 3Al–MoS$_2$ with adsorbed HCN. Red and green regions indicate charge accumulation and depletion, respectively. The isosurface value is set at 0.001~$e/\text{\AA}^3$.}
    \label{fig:2.3}
\end{figure}

Next, the electronic properties and other associated parameters are calculated, which have been summarized in Table~\ref{tab:electronic-parameters}. Taking the recovery time into account, we have provided the band structure (Figure~\ref{fig:2.2}), density of states (DOS), and charge density difference (CDD) plots (Figure~\ref{fig:2.3}) for 3Al–MoS$_2$ only. Other systems have been discussed in this section; however, for detailed structural information, one should refer to the supporting information.

The bandgap of 2P–MoS$_2$ increases by 47.51\%, and that of 3P–MoS$_2$ increases by 43.99\% after the adsorption of the HCN molecule. Similarly, in the case of Si doping, a 12.55\% increase is observed for two-atom doping, whereas for three-atom doping, it decreases by 3.69\% after adsorption. A similar trend of decreasing percentage change in the bandgap is observed in both systems; however, for 2Al–MoS$_2$ and 3Al–MoS$_2$, a 44.43\% rise and a 3.39\% reduction are observed, respectively. The charge transfer ($\Delta Q$) values, work function before HCN adsorption ($\phi_1$), work function after HCN adsorption ($\phi_2$), and change in work function ($\Delta \phi$) are also summarized in Table~\ref{tab:electronic-parameters}. 

Considering all the parameters, the 3Al–MoS$_2$ monolayer emerges as the best candidate among all the two- and three-atom doped monolayers in this study. Despite showing only a 3.39\% change in the band structure, it can still serve as a promising material for designing an HCN gas sensor based on other properties. This small change in the band structure arises from the weak physisorption process; however, it is not too weak, as other properties such as charge transfer and change in work function exhibit significant variations. 

To further confirm its sensing potential, the electrical conductivity and electron emission current density were also calculated at room temperature. Since the bandgap increases after HCN adsorption, the electrical conductivity of the 3Al–MoS$_2$ monolayer decreases by approximately 11.8\% at 300~K. The work function changes by 0.174~eV, which lies within a detectable range. Moreover, when the electron emission current density is calculated using the Richardson equation at room temperature, it is found to increase by nearly 842 times after HCN adsorption.

\begin{figure}[ht]
    \centering
    \includegraphics[width=1.0\textwidth]{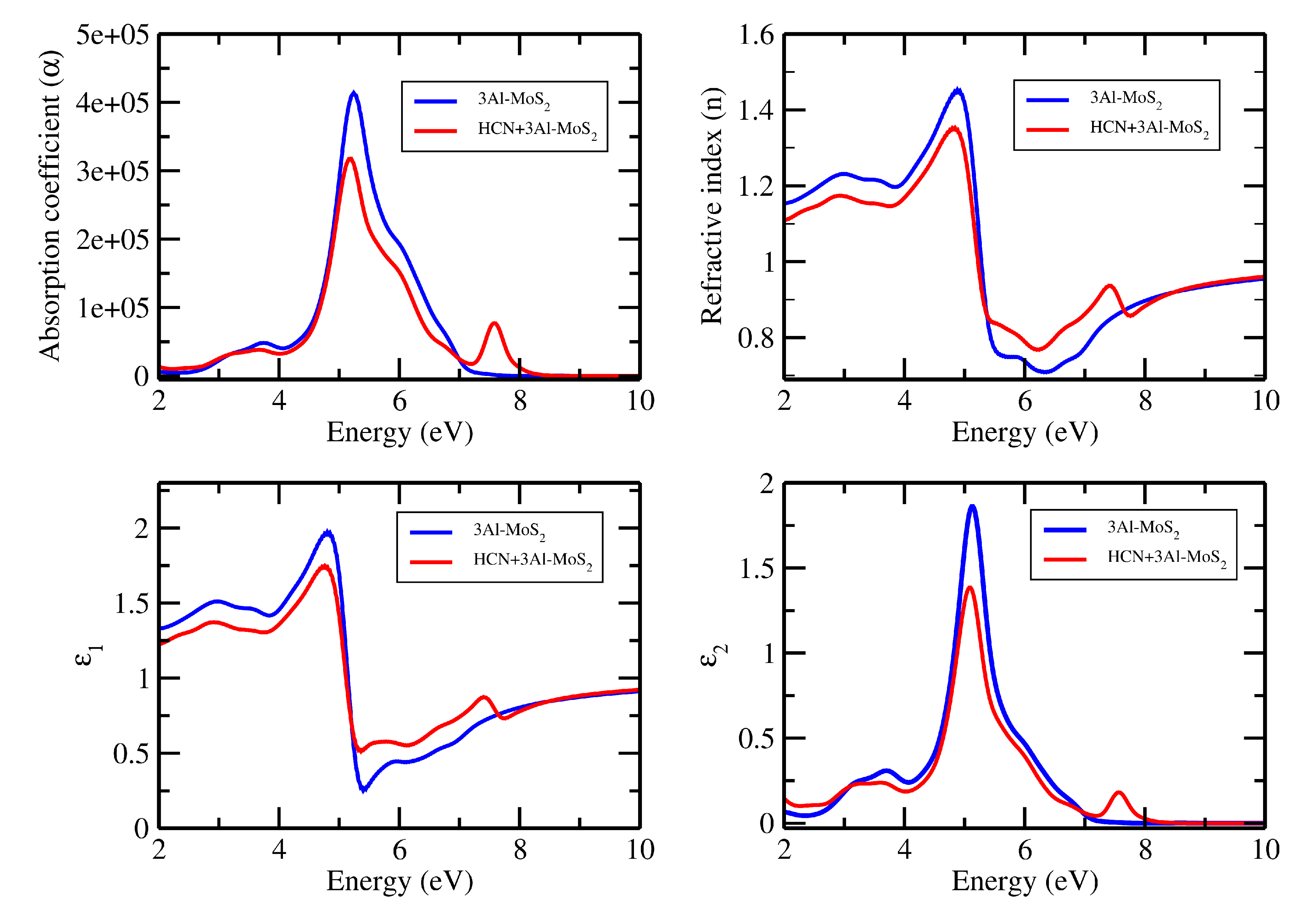}
\caption{Optical properties of pristine 3Al–MoS$_2$ monolayer (blue) and HCN-adsorbed 3Al–MoS$_2$ monolayer (red).}
    \label{fig:optical_Al}
\end{figure}

Finally, we also calculated the optical properties of 3Al–MoS$_2$. The analysis of the optical response of the 3Al–MoS$_2$ monolayer revealed notable modifications upon HCN adsorption, as illustrated in Figure~\ref{fig:optical_Al}. The absorption coefficient decreased by 23.27\% after adsorption. Similarly, reductions of 6.85\%, 11.40\%, and 25.61\% were observed in the refractive index and in the real ($\epsilon_1$) and imaginary ($\epsilon_2$) components of the dielectric function, respectively. The prominent peaks in these optical spectra were located within the photon energy range of 4.80~eV to 5.30~eV, corresponding to the ultraviolet (UV) region of the electromagnetic spectrum. Furthermore, all peaks exhibited a red shift after adsorption, indicating a slight displacement toward lower photon energies.

\section{Conclusion}

This study employs density functional theory (DFT) to evaluate pristine and doped MoS$_2$ monolayers (X–MoS$_2$, where X = P, N, Si, Al, B, Cl) as potential sensors for highly toxic hydrogen cyanide (HCN) gas. Structural optimization confirms good agreement with existing theoretical and experimental data. Adsorption analysis shows that HCN interacts mainly through physisorption with most systems, while Al–MoS$_2$ exhibits the strongest and most stable adsorption. However, recovery-time analysis reveals that Al–MoS$_2$ is impractical for real-time sensing due to an extremely long recovery time (years) at 300~K, whereas Si–MoS$_2$ shows ultrafast recovery in the microsecond range and other systems fall in the nanosecond range.

The analysis of electronic and optical properties before and after HCN adsorption indicates notable changes in band gap, conductivity, work function, and electron emission current density, particularly for P–, Si - and Al–MoS$_2$. When recovery time is considered alongside sensitivity, P–MoS$_2$ and Si–MoS$_2$ emerge as the most promising chemisorption HCN sensors. Additionally, significant variations in optical properties in the ultraviolet (UV) region suggest their applicability in UV-range optical sensing.

The second phase investigates the effect of dopant concentration by substituting two or three sulfur atoms with P, Si, or Al. The substitution of two sulfur atoms enhances the strength of the adsorption, while the addition of a third dopant at a symmetric site weakens the adsorption. Because recovery time depends exponentially on the adsorption energy, controlled doping enables tuning of the sensor performance. Notably, the recovery time increases from nanoseconds to microseconds in 2P–MoS$_2$, while it decreases to about 1.74 seconds in 3Al–MoS$_2$. Based on adsorption behavior, recovery time, and changes in electronic and optical properties, 2P–MoS$_2$ and 3Al–MoS$_2$ are identified as the most efficient candidates for electrochemical and chemiresistive HCN gas sensing, with additional potential for UV-range optical sensor applications.

\section*{Data Availability}
The data supporting the findings of this study are available from the corresponding author upon reasonable request.

\section*{Acknowledgements}
Authors acknowledge the SIESTA team for providing the code. Authors are also thankful to the Department of Physics, HPU Shimla, for providing access to the high-performance computing (HPC) facility. N.T. also sincerely thanks his parents for their constant encouragement, moral and financial support, and motivation throughout this work.

\section*{Author Contributions}
N.T. carried out all calculations, data generation, and analysis, and drafted the original manuscript. A.B. contributed to data analysis and interpretation and co-wrote the original draft. A.K. contributed to data analysis and interpretation, provided technical guidance, and assisted with review and editing. A.S. contributed to data analysis, manuscript writing, review and editing, supervision, coordination, and overall guidance.

\section*{Competing Interests}
The authors declare no competing interests.
\bibliographystyle{unsrt}
\bibliography{references}

\end{document}